# Quantum Kaleidoscopes and Bell's theorem

P.K.Aravind
Physics Department
Worcester Polytechnic Institute
Worcester, MA 01609

ABSTRACT

A quantum kaleidoscope is defined as a set of observables, or states, consisting of many different subsets that provide closely related proofs of the Bell-Kochen-Specker (BKS) and Bell nonlocality theorems. The kaleidoscopes prove the BKS theorem through a simple parity argument, which also doubles as a proof of Bell's nonlocality theorem if use is made of the right sort of entanglement. Three closely related kaleidoscopes are introduced and discussed in this paper: a 15-observable kaleidoscope, a 24-state kaleidoscope and a 60-state kaleidoscope. The close relationship of these kaleidoscopes to a configuration of 12 points and 16 lines known as Reye's configuration is pointed out. The "rotations" needed to make each kaleidoscope yield all its apparitions are laid out. The 60-state kaleidoscope, whose underlying geometrical structure is that of ten interlinked Reyes' configurations (together with their duals), possesses a total of 1120 apparitions that provide proofs of the two Bell theorems. Some applications of these kaleidoscopes to problems in quantum tomography and quantum state estimation are discussed.

# 1. Introduction

By a "quantum kaleidoscope" I mean a set of observables, or states, consisting of many different subsets that prove Bell's theorem [1], so that the effect of shifting one's gaze from one subset to another is somewhat like rotating a kaleidoscope and seeing one beautiful pattern glide into another. The proofs of Bell's theorem yielded by a quantum kaleidoscope are all of a visually obvious kind that requires no more than a simple parity check. The parity check in fact establishes the Bell-Kochen-Specker (BKS) theorem [2,3], but, in conjunction with a suitable entangled state, it serves to establish Bell's nonlocality theorem [1] as well.

This paper presents three different, but related, quantum kaleidoscopes. The first is an "observables" kaleidoscope based on 15 observables for a pair of qubits (or two-state quantum systems) and yields ten slightly different versions of a proof of the BKS theorem first given by Mermin [4]. While the existence of this kaleidoscope was hinted at by Mermin, it came fully into view as a result of an elegant construction by Goodmanson [5]. The author's interest in this kaleidoscope was sparked by the realization that it could be used to prove not only the BKS theorem but Bell's nonlocality theorem as well [6] (a similar point was also made by Cabello [7]).

The second kaleidoscope is a "states" kaleidoscope based on 24 states derived from nine of the fifteen observables above, the states being the simultaneous eigenstates of the six complete sets of commuting observables yielded by these observables. These 24 states were first used by Peres [8] to give a proof of the BKS theorem, but it was later realized [9,10,11] that a host of elegant parity proofs, all differing slightly from each other, could be obtained by choosing different subsets of these 24 states. In fact, the author [11] showed how this 24-state kaleidoscope could be made to yield 16+96 = 112 variations of the two parity proofs of the BKS theorem given in Refs.[9] and [10].

The third kaleidoscope is a 60-state kaleidoscope whose states are the simultaneous eigenstates of the 15 complete sets of commuting observables yielded by the 15 observables for a pair of qubits. This kaleidoscope yields 112 x 10 = 1120 variations of the two basic parity proofs of the BKS theorem mentioned above. The 60-state kaleidoscope is the "state" counterpart of the 15-observable kaleidoscope, and it also incorporates ten overlapping 24-state kaleidoscopes within it. It is the goal of this paper to explain concisely, but fully, the structure of all three kaleidoscopes, to clarify their relationship with each other, and finally to explain how each one can be made to yield all its apparitions.

This paper is organized as follows. Section 2 reviews Goodmanson's construction of the 15-observable kaleidoscope based on the complete graph on a hexagon. Section 3 reviews the author's construction [11] of the 24-state kaleidoscope, in which crucial use is made of a set of 12 points and 16 lines known as Reye's configuration that has been known to projective geometers for over a century. If the states of the 24-state kaleidoscope are regarded as "points" that lie on suitably defined "lines", it turns out that this kaleidoscope can be regarded as a pair of mutually dual Reye's configurations, and that this geometric fact can be used to tease out all its apparitions. What this all means is fully explained. Section 4 builds upon the work of the two preceding sections to give a construction of the 60-state kaleidoscope. The geometric structure underlying this kaleidoscope turns out to be a framework of ten interlinked Reye's configurations, together with their duals.

Aside from providing transparent and closely related proofs of the BKS and Bell theorems, the above kaleidoscopes have some other points of interest. They give rise to "quantum block designs" (our term) that are generalizations of the balanced incomplete block designs known in the field of combinatorics, and they also have some applications to quantum tomography and quantum state estimation. These matters are discussed in the concluding Sec 5.

## 2. 15-observable kaleidoscope

Figure 1 shows the six Pauli operators for a pair of qubits arranged at the vertices of a hexagon. The fifteen edges of the complete graph on this hexagon define fifteen two-qubit observables, each of which is the product of the two single-qubit observables at the extremities of an edge. It is obvious that any two of these observables either commute or anticommute. Goodmanson [5] came up with a simple prescription, based on Fig.1, for telling which is the case: he observed that if the edges corresponding to two observables have no vertices in common they commute, whereas if they have a vertex in common they anticommute. It follows from this that the maximum number of mutually commuting observables is three, since the vertices of the hexagon are then all taken up. It is also easy to see that there are exactly fifteen ways of picking three edges that have no vertices in common, showing that there are exactly fifteen triads of mutually commuting observables for a pair of qubits.

Goodmanson showed how to pick nine of the fifteen observables and arrange them in the form of a 3 x 3 array (a "magic square") that yields a proof of the BKS theorem. His prescription for doing so is the following:

*Goodmanson's construction*: *Form two triangles out of the edges of the complete graph on the hexagon of Fig.1 in such a way that the triangles have no vertices in common. Then arrange the nine observables not included among the edges of these triangles in a 3 x 3 square in such a way that the observables in any row or column form a mutually commuting set (this can be done in only one way up to exchanges of the rows and/ or columns of the square). The resulting "magic square" furnishes a proof of the BKS theorem.*

Figure 2 illustrates the construction of a "magic square" based on the above rule, and Fig.3 shows all the ten squares that can be constructed in this fashion. The ten magic squares of Fig.3 constitute all the different apparitions of this 15-observable kaleidoscope, and they all provide parity proofs of the BKS theorem. (The adjective "magic" is added to square to indicate that the observables in the square lead to a proof of the BKS theorem. A "nonmagic" square is one whose rows and columns consist of commuting observables but that does not lead to a proof of the BKS theorem. It is an interesting fact that Goodmanson's construction, when applied to Fig.1, leads only to the magic squares of Fig.3 and not to any nonmagic squares).

How the magic squares of Fig.3 prove the BKS theorem can be understood as follows [4]. Consider the magic square of Fig.2 and note that the product of the observables in any row or column is $I$ (the identity operator), except for the last column for which the product is $-I$. It follows that if a measurement is made of the (commuting) observables in any row or column of the square, the product of the observed eigenvalues must be +1 (for the observables in the three rows or the first two columns) or -1 (for the observables in the last column). If one now adopts the "realist" position that the eigenvalues reflect properties of the observables that exist prior to measurement, one is faced with the task of assigning +1s and -1s to the nine observables in such a way that the product constraints on the eigenvalues are always met. But the product constraint can be recast as the "sum" constraint that the sum of the eigenvalues in any of the rows or the first two columns be +1 and the sum of the eigenvalues in the last column be -1. However this sum constraint can never be met because, on the one hand, it requires the total number of -1s in the array to be even (if one sums the -1s by the rows) and, on the other hand, it also requires it to be odd (if one sums the -1s by the columns). This contradiction shows the "realist" assumption of preexisting (eigen-)values for the observables to be untenable, and this constitutes Mermin's celebrated proof [4] of the BKS theorem. A similar argument can be connnstructed for the other magic squares in Fig.3.

The above demonstration falls short of proving Bell's nonlocality theorem because of the assumption of "noncontextuality" that went into it, i.e., the assumption that the preexisting value possessed by an observable is independent of whether it is measured as part of a row or column of (commuting) observables. However Ref.[11] shows how an entangled state of four qubits can be used to convert the above BKS proof into a proof of Bell's nonlocality theorem. Thus the ten apparitions of the 15-observable kaleidoscope shown in Fig.3 provide us with joint proofs of both the BKS and Bell theorems, it being understood that a suitable entangled state is used for the latter purpose. It might be added that the two other kaleidoscopes to be discussed in this paper also provide proofs of both the BKS and Bell theorems through a similar stratagem. It is also worth remarking that *any* proof of the BKS theorem can always be converted into a proof of Bell's nonlocality theorem by making use of the right sort of entanglement [12].

## 3. The 24-state kaleidoscope

The first six rows of Fig.4 show the simultaneous eigenstates of the complete sets of commuting observables that make up the rows and columns of the magic square in Fig.2. The 24 states thus obtained form 24 tetrads of mutually orthogonal states that are shown in Fig.5(a); six of these tetrads are just the ones shown in Fig.4, while the remaining 18 arise as mixtures of states from these six tetrads.

To exhibit the kaleidoscope formed by these 24 states, we first introduce the notion of a *Reye's configuration* [13], which is a set of twelve "points" and sixteen "lines" with the property that each "line" passes through exactly three "points" and each "point" has exactly four "lines" passing through it (the quotes around "points" and "lines" are put to indicate that they need have no resemblance to ordinary points and lines and need only satisfy the incidence relations stated). Let us regard each of our 24 states as a "point", with the coordinates of any "point" being the four numbers assigned to that state in Fig.4. Let us also regard two "points" as identical if the coordinates of one are a (possibly complex) multiple of those of the other. Next, let us define a "line" as consisting of all "points" whose coordinates can be expressed as linear combinations, with suitable complex coefficients, of the coordinates of any two "points" on the line. With these definitions of "point" and "line", it is easy to verify that the twelve "points" corresponding to states 1-12 of Fig.4 form a Reye's configuration, with the sixteen "lines" of the configuration being the ones shown to the left of the double arrows in Fig.5(b). A glance at Fig.5(b) shows that each "line" has exactly three "points" on it and each "point" has exactly four "lines" passing through it, as required. It can be verified in a similar manner that the twelve "points" corresponding to states 13-24 of Fig.4 also form a Reye's configuration, with the sixteen lines of this configuration being indicated to the right of the double arrows in Fig.5(b).

The two Reye's configurations introduced above are said to be the "duals" of each other and have the important property that their "lines" can be paired up in such a way that any "point" on a line of one is "orthogonal" to every "point" on the corresponding line of the other (in saying that two "points" are "orthogonal", we mean simply that the corresponding quantum states are orthogonal). We will sometimes speak of the corresponding "lines" in the two configurations as the "partners" of each other. The partner "lines" in the two configurations introduced above are paired up at the ends of the double arrows in Fig.5(b).

The 24-state kaleidoscope has two different types of apparitions: it has sixteen 18-state apparitions and ninety-six 20-state apparitions. The 18-state apparitions can be generated by picking a pair of partner "lines" from the two configurations (which can be done in 16 ways) and retaining only the tetrads in Fig.5(a) that do not contain any of the "points" on the chosen "lines"; this yields exactly nine tetrads, containing 18 states occurring twice each, that furnish a parity proof of the BKS theorem. Three of these 18-state apparitions are shown in Fig.6, with the relevant partner "lines" alongside. The proof of the BKS theorem based on an 18-state apparition involves showing [8,10] that it is impossible to color each of the 18 states red or green in such a way that each of the nine tetrads has exactly one green state in it. But this impossibility is easily demonstrated by noting that, on the one hand, the total number of green states in all the tetrads is required to be odd (because there must be exactly one green state per tetrad) whereas, on the other, it is also required to be even (because each green state is repeated twice over the tetrads). As mentioned earlier, this BKS proof can be extended to a proof of Bell's theorem by making use of a suitable entangled state (in fact, exactly the same state as was used for the 15-observable kaleidoscope).

In order to introduce the 20-state apparitions, it is necessary to define the notion of a "triangle". A "triangle" consists of any three "points" from one of our Reye's configurations such that the inner product of any pair of states corresponding to the "points" has an absolute magnitude of ½. With this definition, the procedure for generating the 20-state apparitions is as follows: pick any "point" from one configuration and any "triangle" from the other in such a way that the "point" from the first is orthogonal to each of the "points" in the "triangle" from the other, and retain only the tetrads in Fig.5(a) that do not contain any of these four "points". This always leads to eleven tetrads, in which eighteen "points" occur twice and two "points" four times, that lead to parity proofs of the BKS theorem. The proof [9] again follows from the fact the total number of tetrads is odd whereas the number of occurrences of each state in them is even, thus making it impossible to color the 20 states in such a way that there is one, and only one, green state in each tetrad. Fig.7(a) lists all the "point"-"triangle" combinations from the two configurations that give rise to the 96 apparitions of this type, and Fig.7(b) shows one such apparition explicitly.

## 4. The 60-state kaleidoscope

The 60 states of this kaleidoscope are the simultaneous eigenstates of the 15 complete sets of commuting observables for a pair of qubits. They are listed in Fig.4 and numbered from 1 to 60. The geometrical structure underlying this kaleidoscope is most easily grasped by looking at Fig.3, which shows the 10 apparitions of the 15-observable kaleidoscope. Each of the "magic squares" in Fig.3 has exactly one row and one column in common with any other square and, consequently, the 24 states arising from any square have exactly 8 states in common with any other square (namely, the states arising as eigenstates of the common row and column). Since any magic square is equivalent to a pair of mutually dual Reye's configurations or to a 24-state kaleidoscope, the 60-state kaleidoscope can be viewed as ten interlocking pieces of either of these types. For example, the 60-state kaleidoscope can be viewed as a set of ten interlocking 24-state kaleidoscopes, with any two of these kaleidoscopes being "welded together" at their common set of 8 states.

The 60-state kaleidoscope has 105 tetrads of mutually orthogonal states, which are shown in Fig.8. As just mentioned, the pattern in Fig.8 can be considered as arising from a fusion of ten 24-state kaleidoscopes of the form shown in Fig.5(a), with any two kaleidoscopes possessing a pair of tetrads in common. Any 24-state kaleidoscope in Fig.8 can be turned into any other one by suitably relabeling the 16 states that change in going from one to the other. The relabelings that cause the 24-state kaleidoscope of Fig.5(a) to pass into the nine kaleidoscopes of this type are shown in Fig.9.

The relabelings of Fig.9 can alternatively be described as unitary transformations that act to transform any magic square in Fig.3 into any other one. The transformations that take the first square in Fig.3 into the remaining squares are shown in Fig.10. Note that all these transformations, with the exception of one, are the products of rotations in the spaces of the individual qubits. Only the transformation between the squares S1 to S6 involves a joint operation in the space of both the qubits; as indicated by the question marks, this transformation has not been identified explicitly yet.

The 60 states of this kaleidoscope have been discussed recently from an alternative point of view by Rigetti et.al. [21]. These authors explored the discretization of Hilbert space with a view to constructing finite sets of states, as well as unitary operators connecting them (gates), that could be made the basis of a "digital" approach to quantum information processing.

## 5. Block designs, tomography and state estimation

The structure underlying the various kaleidoscopes introduced in this paper can be further clarified by recalling the notion of a balanced incomplete block design (BIBD) from the field of combinatorics [14]. A BIBD is a collection of v objects (or "points") consisting of b k-subsets (or "blocks") such that (a) each point belongs to exactly r blocks and (b) any two points occur together exactly λ times in the b blocks. A BIBD is usually denoted by the symbol {b,v,r,k,λ} listing its parameters. The parameters of a BIBD are not independent but satisfy the two constraints bk = vr and r(k-1) = λ(v-1), both of which follow from its definition by simple counting arguments. BIBDs have numerous applications ranging from the statistical design of experiments to the construction of error correcting codes in cryptography.

Let us introduce the notion of a "quantum block design" (QBD) as a BIBD in which quantum states (or observables) play the role of points and orthonormal bases of states (or complete sets of commuting observables) play the role of blocks, but with some additional constraints imposed. To lay out the idea, first consider a QBD in which observables play the role of points and complete sets of commuting observables the role of blocks. A QBD with the symbol $\{b,v,r,k;(\lambda_1,x_1),(\lambda_2,x_2),\ldots,(\lambda_n,x_n)\}$ is defined as set of v observables that form b complete sets of commuting observables, of k observables each, such that (a) each observable occurs in exactly r commuting sets (or "blocks") and (b) each observable occurs $\lambda_1$ times with $x_1$ others, $\lambda_2$ times with $x_2$ others, … and $\lambda_n$ times with $x_n$ others in the commuting subsets (or "blocks"). The parameters of the QBD satisfy the constraints bk = vr and r(k-1) = $\lambda_1 x_1 + \lambda_2 x_2 + \ldots + \lambda_n x_n$ as a consequence of its definition. This definition can be modified in an obvious way to cover the case where states play the role of points and orthonormal bases of states the role of blocks. A QBD reduces to a BIBD if $x_1$ = v-1 and $x_2 = \ldots = x_n = 0$.

With this definition, all the kaleidoscopes introduced earlier can be described as QBDs. The 15-observable, 24-state and 60-state kaleidoscopes can be described as QBDs with the symbols {15,15,3,3;(1,6)}, {24,24,4,4;(1,6),(2,3)} and {105,60,7,4;(1,12),(3,3)}, respectively. The notion of a QBD thus helps to convey many of the structural features of a quantum kaleidoscope in a very concise fashion.

An interesting aspect of the above kaleidoscopes is the way in which they incorporate mutually unbiased bases (MUB) of states or observables within them. Two orthonormal bases $|\alpha_i\rangle$ and $|\beta_i\rangle$ in a $d$-dimensional Hilbert space are said to be mutually unbiased if $|\langle \alpha_i | \beta_j \rangle|^2 = 1/d$ for all $i,j = 1,..,d$. Two complete sets of commuting observables are said to be mutually unbiased (and we will use the term MUB for them as well) if the eigenbases they give rise to are mutually unbiased. MUB play a fundamental role in the quantum theory and also have applications in quantum tomography [15], quantum cryptography [16] and quantum state retrodiction [17]. The observables in any row of a magic square are mutually unbiased with respect to those in any other row, and the same is also true of the columns; however none of the row sets are unbiased with respect to any of the column sets. The maximum number of MUB in a four-dimensional Hilbert space is 5. The 15 triads of commuting observables for a pair of qubits can be partitioned into 6 distinct maximal sets of MUB, which are:

$$\{\sigma_z^1,\sigma_z^2,\sigma_z^1\sigma_z^2\},\{\sigma_x^1,\sigma_x^2,\sigma_x^1\sigma_x^2\},\{\sigma_y^1,\sigma_y^2,\sigma_y^1\sigma_y^2\},\{\sigma_x^1\sigma_y^2,\sigma_y^1\sigma_z^2,\sigma_z^1\sigma_x^2\},\{\sigma_y^1\sigma_x^2,\sigma_z^1\sigma_y^2,\sigma_x^1\sigma_z^2\}$$

$$\{\sigma_z^1,\sigma_z^2,\sigma_z^1\sigma_z^2\},\{\sigma_x^1,\sigma_x^2,\sigma_x^1\sigma_x^2\},\{\sigma_y^1,\sigma_x^2,\sigma_y^1\sigma_x^2\},\{\sigma_z^1\sigma_x^2,\sigma_x^1\sigma_z^2,\sigma_y^1\sigma_y^2\},\{\sigma_y^1\sigma_z^2,\sigma_z^1\sigma_y^2,\sigma_x^1\sigma_x^2\}$$

$$\{\sigma_x^1,\sigma_z^2,\sigma_x^1\sigma_z^2\},\{\sigma_z^1,\sigma_x^2,\sigma_z^1\sigma_x^2\},\{\sigma_y^1,\sigma_y^2,\sigma_y^1\sigma_y^2\},\{\sigma_x^1\sigma_y^2,\sigma_y^1\sigma_x^2,\sigma_z^1\sigma_z^2\},\{\sigma_y^1\sigma_z^2,\sigma_z^1\sigma_y^2,\sigma_x^1\sigma_x^2\}$$ (1)

$$\{\sigma_x^1,\sigma_z^2,\sigma_x^1\sigma_z^2\},\{\sigma_z^1,\sigma_y^2,\sigma_z^1\sigma_y^2\},\{\sigma_y^1,\sigma_x^2,\sigma_y^1\sigma_x^2\},\{\sigma_x^1\sigma_x^2,\sigma_y^1\sigma_y^2,\sigma_z^1\sigma_z^2\},\{\sigma_x^1\sigma_y^2,\sigma_y^1\sigma_z^2,\sigma_z^1\sigma_x^2\}$$

$$\{\sigma_z^1,\sigma_x^2,\sigma_z^1\sigma_x^2\},\{\sigma_y^1,\sigma_z^2,\sigma_y^1\sigma_z^2\},\{\sigma_x^1,\sigma_y^2,\sigma_x^1\sigma_y^2\},\{\sigma_x^1\sigma_x^2,\sigma_y^1\sigma_y^2,\sigma_z^1\sigma_z^2\},\{\sigma_y^1\sigma_x^2,\sigma_z^1\sigma_y^2,\sigma_x^1\sigma_z^2\}$$

$$\{\sigma_x^1,\sigma_x^2,\sigma_x^1\sigma_x^2\},\{\sigma_y^1,\sigma_z^2,\sigma_y^1\sigma_z^2\},\{\sigma_z^1,\sigma_y^2,\sigma_z^1\sigma_y^2\},\{\sigma_x^1\sigma_y^2,\sigma_y^1\sigma_x^2,\sigma_z^1\sigma_z^2\},\{\sigma_z^1\sigma_x^2,\sigma_x^1\sigma_z^2,\sigma_y^1\sigma_y^2\}$$

If we let commuting triads of observables play the role of points and maximal sets of MUB the role of blocks, the above array can be described as a QBD with the symbol {6,15,2,5;(1,8)}. The different maximal sets listed above are actually unitarily equivalent to each other, since any one of them may be turned into any other by a product of suitable single qubit rotations.

The QBD (1) has an interesting application to quantum tomography. In quantum tomography, or "state reconstruction", the object is to determine an unknown quantum state, of which many copies are available, through suitable measurements on the copies. One method of solving this problem is to carry out a measurements of a maximal set of MUBs on the copies, since this yields enough information to allow all the parameters of the unknown density matrix to be determined. Moreover, it was shown by Wootters and Fields [18] that the strategy of measuring MUBs allows the statistical uncertainties in the result to be made as small as possible. To determine an unknown state (pure or mixed) of a pair of qubits, one can therefore carry out a measurement of any one of the six maximal sets of MUBs in (1). This yields all the information needed to determine the 15 parameters of the unknown two-qubit density matrix. However the QBD (1) suggests the following improved strategy: instead of measuring just the 5 triads of observables in an MUB, one measures all 15 triads of observables for the two-qubit system. This increases the number of measurements by a factor of three but, since the 15 triads form six maximal sets of MUB, the number of state determinations goes up by a factor of six, leading to a two to one advantage in terms of data over a strategy based on a single maximal set of MUB. This increased volume of data could be used to check the internal consistency of the state reconstruction and thus further reduce the effect of statistical errors.

The 60-state kaleidoscope also has an application to quantum state estimation. Suppose one is given $N = 2$ or $3$ identical copies of an arbitrary pure state of a two-qubit system and asked to determine the state as best as possible. One way of doing this [19] is to make a suitable generalized measurement (a so-called POVM) on all the copies at once, and to use the result to make a judicious guess about the unknown state. The success of this procedure is gauged by the "fidelity", defined as the squared overlap between the unknown state and the guess for it, averaged over all possible occurrences of the input state. It is known that an unknown pure state of a $d-$state quantum system of which $N$ copies are available can be determined with an average fidelity bounded from above by $(N+1)/(N+d)$. However, designing a POVM that achieves this upper bound for an arbitrary $N$ and $D$ is an unsolved problem. In Ref.[20] it was shown that the states of the 60-state kaleidoscope furnish a solution to this problem for the cases $D = 4$ and $N = 2$ or $3$.

The kaleidoscopes discussed in this paper are by no means the only ones. Other kaleidoscopes, and their applications, will be discussed in a future work.

In conclusion we would like to mention an interesting line of work [22], dubbed "pseudo-telepathy" by its inventors, that makes use of "inequality-free" proofs of Bell's theorem of the sort discussed in this paper. The object of this work is to show how two or more parties who share a suitable form of quantum entanglement can perform certain distributed tasks without any need to communicate with each other – a feat that would be impossible with classical resources alone. This line of work exploits quantum entanglement and ideas from computational complexity theory to address interesting issues/problems in game theory and allied areas.

**Acknowledgements.** I would like to thank Vlad Babau for some helpful discussions during the early stages of this work. I would also like to thank Berge Englert for a fruitful observation about the symmetries of the 60-state kaleidoscope.

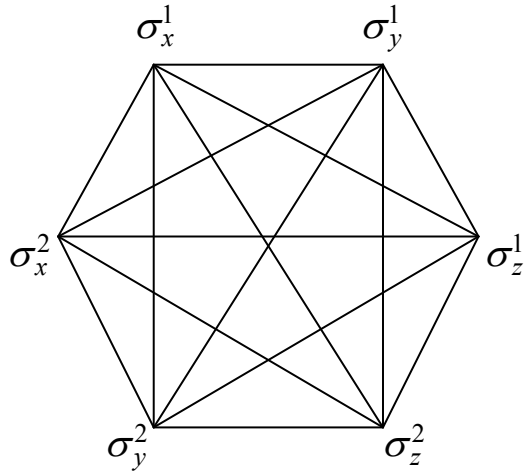

Fig.1. Goodmanson's hexagon. The six Pauli operators for a pair of qubits are arranged at the vertices of a hexagon (with the superscripts 1 and 2 on the operators referring to the qubits). The 15 edges of the complete graph on this hexagon, shown by the lines in the figure, represent 15 observables for a system of two qubits, with each observable being the product of the two single particle observables at the extremities of its edge (with any phase factors of $\pm i$ dropped). For example, the top edge represents the single particle observable $\sigma_z^1$, while the "diameter" joining the top left and bottom right vertices represents the two particle observable $\sigma_x^1 \sigma_z^2$. Two observables whose edges have no vertices in common commute, whereas if they have a vertex in common they anticommute.

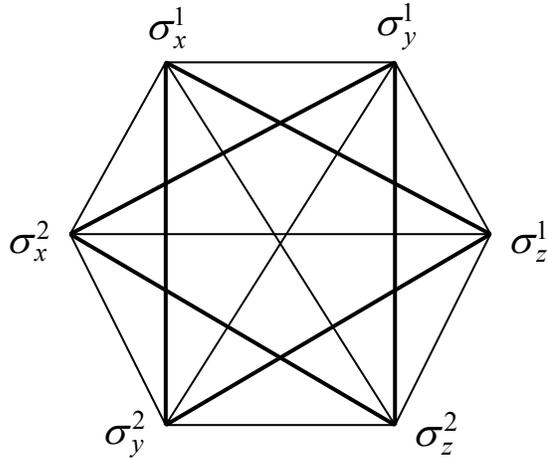

Fig.2. Goodmanson's "magic square" construction. Pick any three vertices of the hexagon and draw in the triangle connecting them, as well as the triangle connecting the remaining three vertices. One example of this is shown at the top, where the two chosen triangles are indicated by the heavy lines. The observables corresponding to the nine remaining edges of the hexagon (shown by the light lines) can be arranged to yield the 3 x 3 "magic square" shown at the bottom, in which each row or column consists of three mutually commuting observables. This magic square, which is unique up to a transposition of its rows and/or columns, can be used to give proofs of both the BKS and Bell theorems (see text for details).

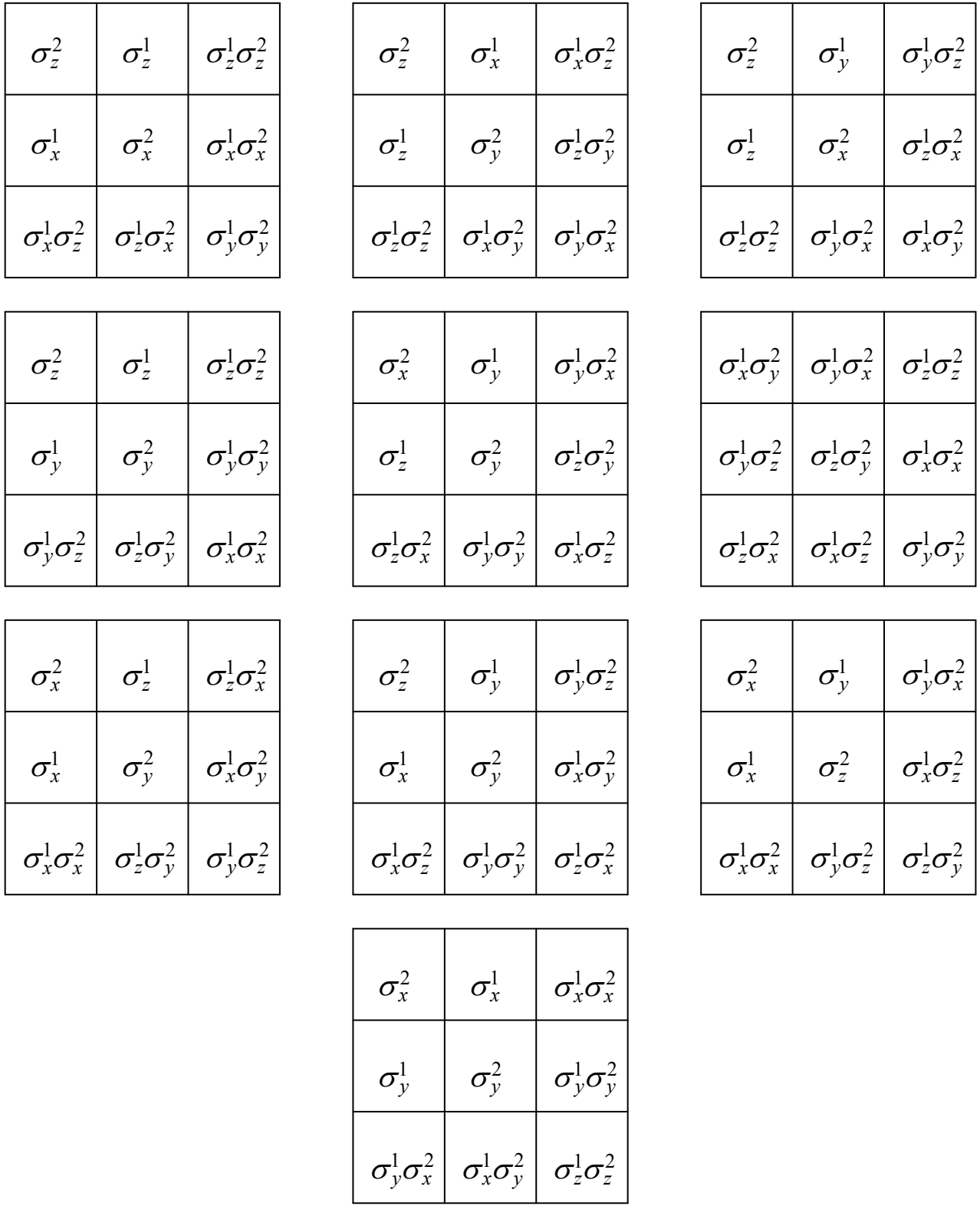

Fig.3. The ten "magic squares" obtained from Fig.1 by Goodmanson's construction.

| Observables | Eigenstates | | | |
|---|---|---|---|---|
| $\sigma_z^1, \sigma_z^2, \sigma_z^1\sigma_z^2$ | 1 = (1,0,0,0) | 2 = (0,1,0,0) | 3 = (0,0,1,0) | 4 = (0,0,0,1) |
| $\sigma_x^1, \sigma_x^2, \sigma_x^1\sigma_x^2$ | 5 = (1,1,1,1) | 6 = (1,-1,1,-1) | 7 = (1,1,-1,-1) | 8 = (1,-1,-1,1) |
| $\sigma_y^1\sigma_y^2, \sigma_x^1\sigma_z^2, \sigma_z^1\sigma_x^2$ | 9 = (1,1,1,-1) | 10 = (1,-1,-1,-1) | 11 = (1,-1,1,1) | 12 = (1,1,-1,1) |
| $\sigma_x^1, \sigma_z^2, \sigma_x^1\sigma_z^2$ | 13 = (1,0,1,0) | 14 = (0,1,0,1) | 15 = (1,0,-1,0) | 16 = (0,1,0,-1) |
| $\sigma_z^1, \sigma_x^2, \sigma_z^1\sigma_x^2$ | 17 = (1,1,0,0) | 18 = (1,-1,0,0) | 19 = (0,0,1,1) | 20 = (0,0,1,-1) |
| * $\sigma_x^1\sigma_x^2, \sigma_z^1\sigma_z^2, \sigma_y^1\sigma_y^2$ | 21 = (1,0,0,1) | 22 = (0,1,1,0) | 23 = (1,0,0,-1) | 24 = (0,1,-1,0) |
| $\sigma_y^1\sigma_z^2, \sigma_z^1\sigma_y^2, \sigma_x^1\sigma_x^2$ | 25 = (1,i,i,1) | 26 = (1,-i,-i,1) | 27 = (1,-i,i,-1) | 28 = (1,i,-i,-1) |
| * $\sigma_x^1\sigma_y^2, \sigma_y^1\sigma_z^2, \sigma_z^1\sigma_x^2$ | 29 = (1,-1,i,i) | 30 = (1,1,-i,i) | 31 = (1,1,i,-i) | 32 = (1,-1,-i,-i) |
| * $\sigma_y^1\sigma_x^2, \sigma_z^1\sigma_y^2, \sigma_x^1\sigma_z^2$ | 33 = (1,i,-1,i) | 34 = (1,-i,1,i) | 35 = (1,i,1,-i) | 36 = (1,-i,-1,-i) |
| $\sigma_x^1\sigma_y^2, \sigma_y^1\sigma_x^2, \sigma_z^1\sigma_z^2$ | 37 = (1,0,0,i) | 38 = (1,0,0,-i) | 39 = (0,1,i,0) | 40 = (0,1,-i,0) |
| $\sigma_y^1, \sigma_z^2, \sigma_y^1\sigma_z^2$ | 41 = (1,0,i,0) | 42 = (1,0,-i,0) | 43 = (0,1,0,i) | 44 = (0,1,0,-i) |
| $\sigma_y^1, \sigma_y^2, \sigma_y^1\sigma_y^2$ | 45 = (1,i,i,-1) | 46 = (1,-i,-i,-1) | 47 = (1,i,-i,1) | 48 = (1,-i,i,1) |
| $\sigma_z^1, \sigma_y^2, \sigma_z^1\sigma_y^2$ | 49 = (1,i,0,0) | 50 = (1,-i,0,0) | 51 = (0,0,1,i) | 52 = (0,0,1,-i) |
| $\sigma_x^1, \sigma_y^2, \sigma_x^1\sigma_y^2$ | 53 = (1,i,1,i) | 54 = (1,-i,1,-i) | 55 = (1,i,-1,-i) | 56 = (1,-i,-1,i) |
| $\sigma_y^1, \sigma_x^2, \sigma_y^1\sigma_x^2$ | 57 = (1,1,i,i) | 58 = (1,-1,i,-i) | 59 = (1,1,-i,-i) | 60 = (1,-1,-i,i) |

Fig.4. Eigenstates of the 15 triads of commuting observables for a system of two qubits. The first column shows the triads of commuting observables and the next four columns the four orthogonal eigenstates arising as simultaneous eigenstates of the observables in each triad. Each (unnormalized) eigenstate has the form $a|00\rangle + b|01\rangle + c|10\rangle + d|11\rangle$ in the standard basis of a pair of qubits, and this is abbreviated to the set of four numbers $(a,b,c,d)$ (with i = √-1). The eigenstates have been given the labels 1 through 60 for ease of reference. Each eigenstate has the eigenvalue +1 or -1 with respect to each of the commuting observables, and the eigenvalue signatures of each eigenstate in a row are unique. The product of the eigenvalues for each of the eigenstates in the three rows marked by a * is -1, whereas it is +1 for the eigenstates in all the remaining rows.

```
 1  2  3  4      3  4 17 18      6  8 17 19     10 11 17 20
 1  2 19 20      5  6  7  8      7  8 13 14     10 12 13 16
 1  3 14 16      5  6 15 16      9 10 11 12     11 12 22 23
 1  4 22 24      5  7 18 20      9 10 21 24     13 14 15 16
 2  3 21 23      5  8 23 24      9 11 14 15     17 18 19 20
 2  4 13 15      6  7 21 22      9 12 18 19     21 22 23 24
```

Fig.5(a). A 24-state kaleidoscope, consisting of 24 tetrads of mutually orthogonal states. The states of this kaleidoscope arise as the joint eigenstates of the observables in a row or column of the "magic square" in Fig.2. These states are listed in the first six rows of Table 4 and they are referred to by their labels there.

```
(1,5,10) ↔ (16,20,24)        (3,7,9)  ↔ (14,18,21)
(1,6,12) ↔ (16,19,22)        (3,6,10) ↔ (16,17,21)
(1,7,11) ↔ (14,20,22)        (3,5,12) ↔ (16,18,23)
(1,8,9)  ↔ (14,19,24)        (3,8,11) ↔ (14,17,23)
(2,7,10) ↔ (13,20,21)        (4,5,9)  ↔ (15,18,24)
(2,5,11) ↔ (15,20,23)        (4,8,10) ↔ (13,17,24)
(2,6,9)  ↔ (15,19,21)        (4,7,12) ↔ (13,18,22)
(2,8,12) ↔ (13,19,23)        (4,6,11) ↔ (15,17,22)
```

Fig.5(b). The 32 "lines" formed by the "points" (i.e. quantum states) 1-24 of Fig.4. The "points" 1-12 form a Reye's configuration, with the 16 "lines" of this configuration being shown to the left of the double arrows (each line is indicated simply by the three "points" that lie on it). Similarly, the "points" 13-24 form another Reye's configuration (dual to the first), whose 16 "lines" are indicated to the right of the double arrows. The paired lines in the two configurations, referred to as the "partners" of each other in the text, have the property that any state in one is orthogonal to all the states in the other.

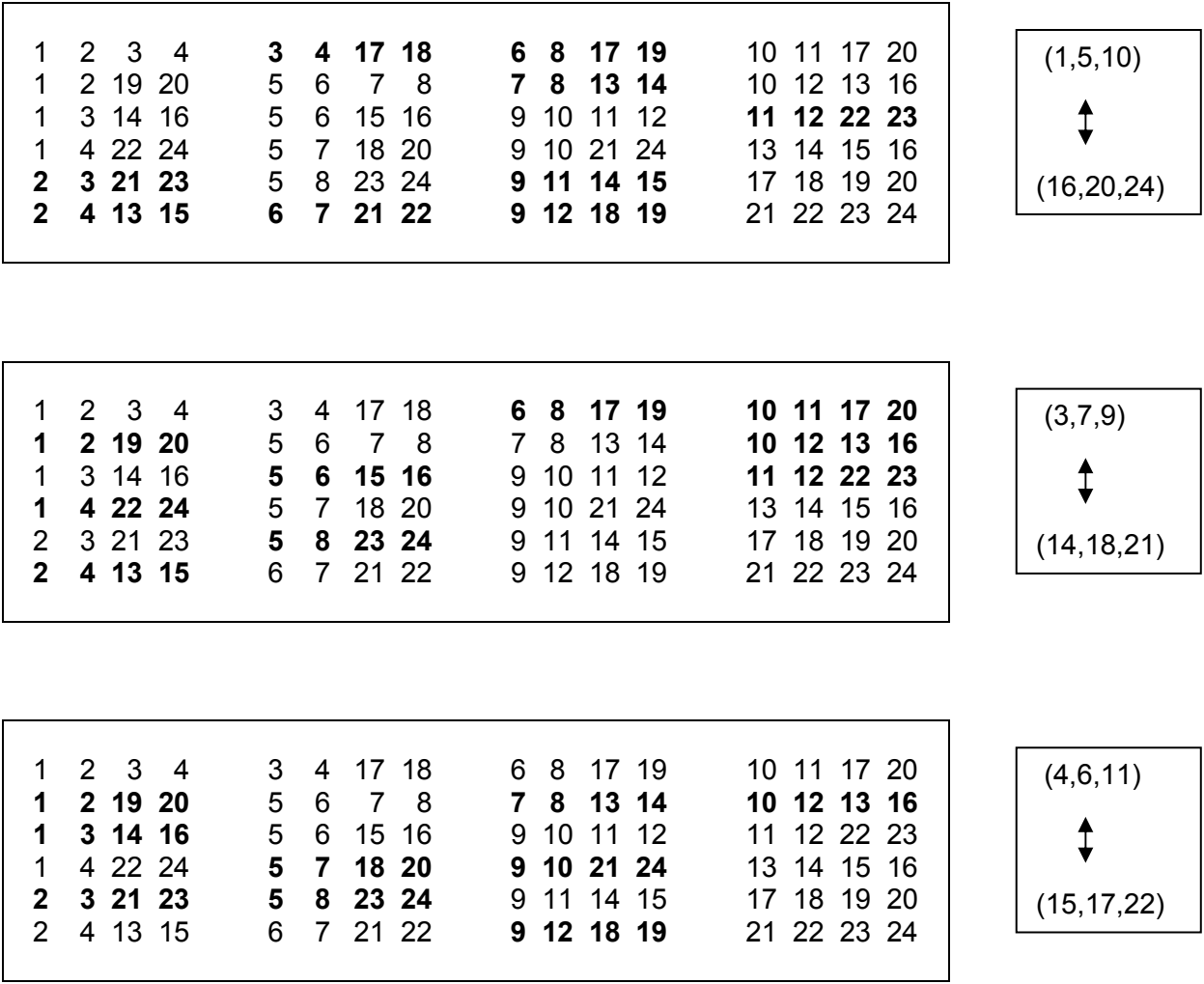

Fig.6. Three 18-state apparitions of the 24-state kaleidoscope of Fig.5(a). Each apparition is obtained by retaining only the tetrads not containing any of the six states in the pair of "partner" lines shown to its right. There are always nine such tetrads, and they are shown in boldface above. These nine tetrads, which contain 18 states each occurring twice, provide a parity proof of the BKS theorem.

```
1  →  {14,19,22},  {14,20,24},  {16,19,24},  {16,20,22}
2  →  {13,19,21},  {13,20,23},  {15,19,23},  {15,20,21}
3  →  {14,17,21},  {14,18,23},  {16,17,23},  {16,18,21}
4  →  {13,17,22},  {13,18,24},  {15,17,24},  {15,18,22}
5  →  {15,18,23},  {15,20,24},  {16,18,24},  {16,20,23}
6  →  {15,17,21},  {15,19,22},  {16,17,22},  {16,19,21}
7  →  {13,18,21},  {13,20,22},  {14,18,22},  {14,20,21}
8  →  {13,17,23},  {13,19,24},  {14,17,24},  {14,19,23}
9  →  {14,18,24},  {14,19,21},  {15,18,21},  {15,19,24}
10 →  {13,17,21},  {13,20,24},  {16,17,24},  {16,20,21}
11 →  {14,17,22},  {14,20,23},  {15,17,23},  {15,20,22}
12 →  {13,18,23},  {13,19,22},  {16,18,22},  {16,19,23}
13 →  {2,7,12},    {2,8,10},    {4,7,10},    {4,8,12}
14 →  {1,7,9},     {1,8,11},    {3,7,11},    {3,8,9}
15 →  {2,5,9},     {2,6,11},    {4,5,11},    {4,6,9}
16 →  {1,5,12},    {1,6,10},    {3,5,10},    {3,6,12}
17 →  {3,6,11},    {3,8,10},    {4,6,10},    {4,8,11}
18 →  {3,5,9},     {3,7,12},    {4,5,12},    {4,7,9}
19 →  {1,6,9},     {1,8,12},    {2,6,12},    {2,8,9}
20 →  {1,5,11},    {1,7,10},    {2,5,10},    {2,7,11}
21 →  {2,6,10},    {2,7,9},     {3,6,9},     {3,7,10}
22 →  {1,6,11},    {1,7,12},    {4,6,12},    {4,7,11}
23 →  {2,5,12},    {2,8,11},    {3,5,11},    {3,8,12}
24 →  {1,5,9},     {1,8,10},    {4,5,10},    {4,8,9}
```

Fig.7(a). How to construct the 20-state apparitions of the 24-state kaleidoscope of Fig.5(a). For any of the states 1 to 24, pick one of the four sets of three states shown to its right and keep only the tetrads not containing any of these four states; the 11 tetrads so obtained provide a parity proof of the BKS theorem.

```
1  2  3   4      3  4  17 18     6  8  17 19     10 11 17 20
1  2  19 20      5  6  7  8      7  8  13 14     10 12 13 16
1  3  14 16      5  6  15 16     9  10 11 12     11 12 22 23
1  4  22 24      5  7  18 20     9  10 21 24     13 14 15 16
2  3  21 23      5  8  23 24     9  11 14 15     17 18 19 20
2  4  13 15      6  7  21 22     9  12 18 19     21 22 23 24
```

Fig.7(b) Illustrating the construction of Fig.7(a) based on selection of the states 1,14,19,22 and the exclusion of all tetrads containing these states. The remaining 11 tetrads, shown in boldface, contain 18 states occurring twice each and 2 states four times each and provide a parity proof of the BKS theorem.

```
 1  2  3  4     6  7 25 26    13 16 33 36    25 28 34 36    37 39 58 59
 1  2 19 20     6  8 17 19    14 15 34 35    25 28 50 51    37 40 54 55
 1  2 51 52     6  8 57 59    14 16 41 42    26 27 33 35    38 39 53 56
 1  3 14 16     7  8 13 14    15 16 53 54    26 27 49 52    38 40 57 60
 1  3 43 44     7  8 53 54    17 18 19 20    26 28 29 31    41 42 43 44
 1  4 22 24     9 10 11 12    17 18 51 52    26 28 41 44    41 43 46 47
 1  4 39 40     9 10 21 24    17 19 58 60    29 30 31 32    41 43 59 60
 2  3 21 23     9 10 47 48    17 20 29 32    29 30 38 39    42 44 45 48
 2  3 37 38     9 11 14 15    18 19 30 31    29 30 54 55    42 44 57 58
 2  4 13 15     9 11 33 36    18 20 57 59    29 31 42 43    45 46 47 48
 2  4 41 42     9 12 18 19    19 20 49 50    30 32 41 44    45 47 50 52
 3  4 17 18     9 12 29 32    21 22 23 24    31 32 37 40    45 47 54 56
 3  4 49 50    10 11 17 20    21 22 27 28    31 32 53 56    45 48 59 60
 5  6  7  8    10 11 30 31    21 23 39 40    33 34 35 36    46 47 57 58
 5  6 15 16    10 12 13 16    21 24 45 46    33 34 38 40    46 48 49 51
 5  6 55 56    10 12 34 35    22 23 47 48    33 34 58 59    46 48 53 55
 5  7 18 20    11 12 22 23    22 24 37 38    33 35 50 51    49 50 51 52
 5  7 58 60    11 12 45 46    23 24 25 26    34 36 49 52    49 51 54 56
 5  8 23 24    13 14 15 16    25 26 27 28    35 36 37 39    50 52 53 55
 5  8 27 28    13 14 55 56    25 27 30 32    35 36 57 60    53 54 55 56
 6  7 21 22    13 15 43 44    25 27 42 43    37 38 39 40    57 58 59 60
```

Fig.8. The 60-state kaleidoscope, made up of the 60 states of Fig.4. This kaleidoscope consists of 105 tetrads of mutually orthogonal states. Each state occurs in seven tetrads, and occurs with three other states thrice each and nine other states once each. This kaleidoscope has 10 partially overlapping 24-state kaleidoscopes embedded within it, as explained in the caption to Fig.9.

| S1 |   | S2 | S3 | S4 | S5 | S6 | S7 | S8 | S9 | S10 |
|----|---|----|----|----|----|----|----|----|----|-----|
| 1  | → | 1  | 1  | 1  | 50 | 37 | 49 | 41 | 42 | 37 |
| 2  | → | 2  | 2  | 2  | 49 | 40 | 50 | 44 | 44 | 39 |
| 3  | → | 3  | 3  | 3  | 51 | 39 | 51 | 42 | 41 | 40 |
| 4  | → | 4  | 4  | 4  | 52 | 38 | 52 | 43 | 43 | 38 |
| 5  | → | 53 | 57 | 25 | 57 | 26 | 5  | 54 | 5  | 5  |
| 6  | → | 54 | 58 | 27 | 60 | 27 | 6  | 53 | 6  | 6  |
| 7  | → | 55 | 59 | 28 | 59 | 28 | 7  | 55 | 7  | 7  |
| 8  | → | 56 | 60 | 26 | 58 | 25 | 8  | 56 | 8  | 8  |
| 9  | → | 35 | 31 | 45 | 9  | 9  | 30 | 9  | 35 | 45 |
| 10 | → | 36 | 32 | 46 | 10 | 10 | 32 | 10 | 33 | 46 |
| 11 | → | 34 | 29 | 48 | 11 | 11 | 29 | 11 | 34 | 47 |
| 12 | → | 33 | 30 | 47 | 12 | 12 | 31 | 12 | 36 | 48 |
| 13 | → | 13 | 41 | 41 | 34 | 34 | 53 | 13 | 13 | 53 |
| 14 | → | 14 | 43 | 44 | 33 | 36 | 54 | 14 | 14 | 54 |
| 15 | → | 15 | 42 | 42 | 36 | 33 | 55 | 15 | 15 | 56 |
| 16 | → | 16 | 44 | 43 | 35 | 35 | 56 | 16 | 16 | 55 |
| 17 | → | 49 | 17 | 49 | 17 | 30 | 17 | 31 | 59 | 57 |
| 18 | → | 50 | 18 | 50 | 18 | 29 | 18 | 29 | 60 | 60 |
| 19 | → | 51 | 19 | 52 | 19 | 32 | 19 | 32 | 57 | 59 |
| 20 | → | 52 | 20 | 51 | 20 | 31 | 20 | 30 | 58 | 58 |
| 21 | → | 37 | 37 | 21 | 48 | 21 | 25 | 48 | 26 | 21 |
| 22 | → | 40 | 39 | 22 | 45 | 22 | 26 | 46 | 25 | 22 |
| 23 | → | 38 | 38 | 23 | 46 | 23 | 28 | 45 | 28 | 23 |
| 24 | → | 39 | 40 | 24 | 47 | 24 | 27 | 47 | 27 | 24 |

Fig.9. The 24-state kaleidoscope of Fig.5(a) can be turned into nine other such kaleidoscopes by replacing the states in it by other states in the manner shown in the above table. If we introduce the labels S1, S2, …, S10 for the ten "magic squares" of Fig.3 from left to right and top to bottom, and use the same labels for the 24-state kaleidoscopes derived from them, the above table shows how the kaleidoscope S1 of Fig.5(a) can be turned into any of the other nine kaleidoscopes. Note that in the transformation from S1 to any of the other kaleidoscopes, 8 of the states remain fixed and only the other 16 states get replaced.

| S1 → S2  | $\sigma_x^1 \leftrightarrow \sigma_z^1$ , $\sigma_x^2 \leftrightarrow \sigma_y^2$ |
|---|---|
| S1 → S3  | $\sigma_x^1 \to \sigma_z^1 \to \sigma_y^1 \to \sigma_x^1$ |
| S1 → S4  | $\sigma_x^1 \leftrightarrow \sigma_y^1$ , $\sigma_x^2 \leftrightarrow \sigma_y^2$ |
| S1 → S5  | $\sigma_x^1 \to \sigma_z^1 \to \sigma_y^1 \to \sigma_x^1$ , $\sigma_x^2 \to \sigma_y^2 \to \sigma_z^2 \to \sigma_x^2$ |
| S1 → S6  | ??? |
| S1 → S7  | $\sigma_x^2 \to \sigma_y^2 \to \sigma_z^2 \to \sigma_x^2$ |
| S1 → S8  | $\sigma_z^1 \leftrightarrow \sigma_y^1$ , $\sigma_x^2 \leftrightarrow \sigma_y^2$ |
| S1 → S9  | $\sigma_y^1 \leftrightarrow \sigma_z^1$ , $\sigma_x^2 \leftrightarrow \sigma_z^2$ |
| S1 → S10 | $\sigma_x^1 \to \sigma_y^1 \to \sigma_z^1 \to \sigma_x^1$ , $\sigma_x^2 \to \sigma_y^2 \to \sigma_z^2 \to \sigma_x^2$ |

Fig.10. The unitary transformations that take the magic square S1 (or the corresponding 24-state kaleidoscope) into the other nine magic squares (or the corresponding kaleidoscopes). The labels S1, S2,… are explained in the caption to Fig.9. All the transformations, with the exception of that from S1 to S6, involve independent rotations on the two qubits. The transformation from S1 to S6 involves a joint rotation in the space of both the qubits and has not been explicitly identified, as indicated by the question marks.